\documentclass[prl,aps,showpacs,amsmath,twocolumn,a4paper,superscriptaddress]{revtex4}
 \pdfoutput=1
\usepackage{bm}
\usepackage{graphicx}
 \usepackage{color}

\begin{document}

\title{Mixing times in evolutionary game dynamics}

\author{Andrew J. Black}
\email{andrew.black@adelaide.edu.au}
\affiliation{School of Mathematical Sciences, The University of Adelaide, Adelaide, SA 5005, Australia}
\affiliation{Theoretical Physics, School of Physics and Astronomy, University of 
Manchester, Manchester M13 9PL, United Kingdom}

\author{Arne Traulsen}
\email{traulsen@evolbio.mpg.de}
\affiliation{Max-Planck-Institute for Evolutionary Biology, August-Thienemann-Str. 2, 24306 Pl\"on, Germany}

\author{Tobias Galla}
\email{tobias.galla@manchester.ac.uk}
\affiliation{Theoretical Physics, School of Physics and Astronomy, University of 
Manchester, Manchester M13 9PL, United Kingdom}

\newcommand{\beq}{\begin{equation}}
\newcommand{\eeq}{\end{equation}}
\newcommand{\bal}{\begin{aligned}}
\newcommand{\eal}{\end{aligned}}

\newcommand{\be}{\begin{equation}}
\newcommand{\ee}{\end{equation}}
\newcommand{\bd}{\begin{displaymath}}
\newcommand{\ed}{\end{displaymath}}
\newcommand{\BE}{\begin{eqnarray}}
\newcommand{\EE}{\end{eqnarray}}
\newcommand{\vsp}{\vspace*{3mm}}
\newcommand{\pprime}{{\prime\prime}}
\newcommand{\R}{{\rm I\!R}}
\newcommand{\order}{{\cal O}}
\newcommand{\smallo}{{o}}
\newcommand{\plus}{{\!+\!}}
\newcommand{\minus}{{\!-\!}}
\newcommand{\sgn}{{\rm sgn}}
\newcommand{\erf}{{\rm erf}}
\newcommand{\erfc}{{\rm erfc}}
\newcommand{\id}{{\openone}}
\newcommand{\real}{\mathbb R}
\newcommand{\ba}{\ensuremath{\mathbf{a}}}
\newcommand{\bb}{\ensuremath{\mathbf{b}}}
\newcommand{\bh}{\ensuremath{\mathbf{h}}}
\newcommand{\bk}{\ensuremath{\mathbf{k}}}
\newcommand{\bq}{\ensuremath{\mathbf{q}}}
\newcommand{\bu}{\ensuremath{\mathbf{u}}}
\newcommand{\bv}{\ensuremath{\mathbf{v}}}
\newcommand{\bw}{\ensuremath{\mathbf{w}}}
\newcommand{\bx}{\ensuremath{\mathbf{x}}}
\newcommand{\by}{\ensuremath{\mathbf{y}}}
\newcommand{\bz}{\ensuremath{\mathbf{z}}}
\newcommand{\bn}{\ensuremath{\mathbf{n}}}
\newcommand{\bolde}{\ensuremath{\mathbf{e}}}

\newcommand{\bA}{\ensuremath{\mathbf{A}}}
\newcommand{\bB}{\ensuremath{\mathbf{B}}}
\newcommand{\bC}{\ensuremath{\mathbf{C}}}
\newcommand{\bD}{\ensuremath{\mathbf{D}}}
\newcommand{\bF}{\ensuremath{\mathbf{F}}}
\newcommand{\bG}{\ensuremath{\mathbf{G}}}
\newcommand{\bGz}{\ensuremath{\mathbf{G_0}}}
\newcommand{\bRone}{\ensuremath{\mathbf{R_1}}}
\newcommand{\bJ}{\ensuremath{\mathbf{J}}}
\newcommand{\bK}{\ensuremath{\mathbf{K}}}
\newcommand{\bR}{\ensuremath{\mathbf{R}}}
\newcommand{\bT}{\ensuremath{\mathbf{T}}}
\newcommand{\bW}{\ensuremath{\mathbf{W}}}
\newcommand{\bM}{\ensuremath{\mathbf{M}}}
\newcommand{\bp}{\ensuremath{\mathbf{p}}}

\newcommand{\bCx}{\ensuremath{\mathbf{C_1}}}
\newcommand{\bCy}{\ensuremath{\mathbf{C_2}}}
\newcommand{\bGx}{\ensuremath{\mathbf{G_1}}}
\newcommand{\bGy}{\ensuremath{\mathbf{G_2}}}

\newcommand{\hq}{\hat{q}}
\newcommand{\hw}{\hat{w}}
\newcommand{\hx}{\hat{x}}
\newcommand{\hy}{\hat{y}}
\newcommand{\hA}{\hat{A}}
\newcommand{\hC}{\hat{C}}
\newcommand{\hG}{\hat{G}}
\newcommand{\hK}{\hat{K}}
\newcommand{\hL}{\hat{L}}
\newcommand{\D}{{\cal D}}
\newcommand{\hbq}{\hat{\mbox{\boldmath $q$}}}
\newcommand{\qbo}{{\mbox{\boldmath $q$}}}
\newcommand{\hbx}{\hat{\mbox{\boldmath $x$}}}
\newcommand{\hbw}{\hat{\mbox{\boldmath $w$}}}
\newcommand{\hOmega}{\hat{\mbox{$\Omega$}}}
\newcommand{\bchi}{{\mbox{\boldmath $\chi$}}}
\newcommand{\bomega}{{\mbox{\boldmath $\omega$}}}
\newcommand{\boldpsi}{{\mbox{\boldmath $\psi$}}}
\newcommand{\boldphi}{{\mbox{\boldmath $\varphi$}}}
\newcommand{\bEta}{{\mbox{\boldmath $\eta$}}}
\newcommand{\bzeta}{{\mbox{\boldmath $\zeta$}}}
\newcommand{\bsigma}{{\mbox{\boldmath $\sigma$}}}
\newcommand{\blambda}{{\mbox{\boldmath $\lambda$}}}
\newcommand{\bmu}{{\mbox{\boldmath $\mu$}}}
\newcommand{\bnu}{{\mbox{\boldmath $\nu$}}}
\newcommand{\btau}{{\mbox{\boldmath $\tau$}}}
\newcommand{\boldOmega}{{\mbox{\boldmath $\Omega$}}}
\newcommand{\bxi}{\bm{\xi}}
\newcommand{\olx}{\overline{\mathbf{x}}}
\newcommand{\olxone}{\overline{x}_1}
\newcommand{\olxtwo}{\overline{x}_2}
\newcommand{\olxonedot}{\dot{\overline{x}}_1}
\newcommand{\olxtwodot}{\dot{\overline{x}}_2}
\newcommand{\bnull}{{\mbox{\boldmath $0$}}}
\newcommand{\rate}{\tilde{\eta}}
\newcommand{\double}{{\prime\prime}}
\newcommand{\tp}{t^\prime}
\newcommand{\td}{t^{\prime\prime}}
\renewcommand{\labelenumi}{(\roman{enumi})}
\newcommand{\wt}{\widetilde}
\newcommand{\avg}[1]{\left\langle{#1}\right\rangle}
\newcommand{\davg}[1]{\left\langle\left\langle{#1}\right\rangle\right\rangle}
\newcommand{\fE}{\mathbb{E}}
\newcommand{\shifta}{\widehat{E}_1} 
\newcommand{\shiftb}{\widehat{E}_2} %
\newcommand{\shifti}{\widehat{E}_i}
\newcommand{\pd}[2]{\frac{\partial #1}{\partial #2}}
\newcommand{\ident}{\mathbb{I}}

\newcommand{\Bra}[1]{\ensuremath{\langle#1|}}
\newcommand{\bra}[1]{\ensuremath{\langle#1|}}
\newcommand{\Ket}[1]{\ensuremath{|#1\rangle}}
\newcommand{\ket}[1]{\ensuremath{|#1\rangle}}
\newcommand{\BraKet}[2]{\ensuremath{\langle #1|#2\rangle}}
\newcommand{\braket}[2]{\ensuremath{\langle #1|#2\rangle}}
\newcommand{\KetBra}[1]{\ensuremath{| #1 \rangle \langle #1 |}}
\newcommand{\ketbra}[1]{\ensuremath{| #1 \rangle \langle #1 |}}
\newcommand{\KetBraO}[3]{\ensuremath{| #1 \rangle_{#3}\langle #2 |}}
\newcommand{\Eins}{\ensuremath{\mathbbm{1}}}
\newcommand{\eins}{\ensuremath{\mathbbm{1}}}
\newcommand{\HH}{\ensuremath{\mathcal{H}}}
\newcommand{\NN}{\ensuremath{\mathcal{N}}}
\newcommand{\LL}{\ensuremath{\mathcal{L}}}
\newcommand{\TT}{\ensuremath{\mathcal{T}}}
\newcommand{\PP}{\ensuremath{\mathcal{P}}}
\newcommand{\QQ}{\ensuremath{\mathcal{Q}}}
\newcommand{\RR}{\ensuremath{\mathcal{R}}}
\newcommand{\EEE}{\ensuremath{\mathcal{E}}}

\date{\today}
\begin{abstract}
 
Without mutation and migration, evolutionary dynamics ultimately leads to the extinction of all but one species. Such fixation processes are well understood and can be characterized analytically with methods from statistical physics. However, many biological arguments focus on stationary distributions in a mutation-selection equilibrium. 
Here, we address the equilibration time required to reach stationarity in the presence of mutation, this is known as the mixing time in the theory of Markov processes. 
We show that mixing times in evolutionary games have the opposite behaviour from fixation times when the intensity of selection increases: 
In coordination games with bistabilities, the fixation time decreases, but the mixing time increases. 
In coexistence games with metastable states, the fixation time increases, but the mixing time decreases. 
Our results are based on simulations and the WKB approximation of the master equation.

\end{abstract}


\maketitle

How long does it take for a stochastic many-particle system to reach its stationary distribution? This question goes beyond traditional equilibrium statistical physics and requires a theory for non-equilibrium systems. Significant progress has been made over the last decades, but developing a more complete theory is still very much work in progress. Many non-equilibrium systems lack an energy or Lyapunov function, any theoretical analysis has to start from the microscopic dynamics itself. 
Such approaches have been applied successfully to off-equilibrium phenomena in physics \cite{Schmittmann}, but also to a number of applications in adjacent disciplines, including in epidemiology, biological transport, pattern formation, agent-based models in economics and of social phenomena \cite{schadschneider, Cross, Fortunato, Helbing}.

For stochastic processes with absorbing states, our opening question can be answered -- at least to some extent.
Absorbing states are those in which the system gets `trapped', so that a full dynamic arrest occurs. 
Systems with absorbing states exhibit new types of phase transitions, universality classes and complexity, previously unknown in physics \cite{Hinrichsen,Henkel}. 
They are relevant in social systems, where an absorbing state may correspond to a uniform consensus, and in evolutionary biology where they describe fixation of a trait. 
Stochasticity can also drive individual phenotypes to extinction in evolutionary game dynamics. 
In the absence of mutation, a given phenotype is never re-introduced once it has become eliminated from the population. 
Answering the question of equilibration times for this type of systems then amounts to calculating the time to fixation \cite{Ewens,Antal}.

The purpose of our work is to develop a similar approach for evolutionary systems with mutation. 
In such systems there are no absorbing states and thus no fixation. Still, they reach a stationary distribution at asymptotic times. 
In order to characterize the approach to stationarity we consider what is referred to as the {\em mixing time} in the theory of Markov process \cite{Levin}. 
This is the time needed for the probability distribution over states to approach its stationary shape up to some specified small distance $\varepsilon$. 
Mixing times have been considered in the context of Markov Chain Monte Carlo methods \cite{Levin}, and recently in game dynamical learning \cite{Auletta}, but, to our knowledge, they have not been discussed for evolutionary processes. We here introduce the basic concepts, analyze mixing times in $2\times 2$ evolutionary games and show how methods from quasi-classical physics can be used to obtain analytical approximations.
Our analysis is based on computer simulations and analytical approximations using the Wentzel-Kramers-Brillouin (WKB) method \cite{Kramers,Bender}. 
While we focus on specific instances of evolutionary dynamics, we expect that these tools 
can be applied to describe the non-equilibrium dynamics of a large class of individual-based models.

We consider a well-mixed population of $N$ individuals of type $A$ or $B$. The state of the system is determined by the number $n\in\{0,\dots,N\}$ of individuals of type $A$. The fitness of individuals of the two types interacting in an evolutionary $2\times 2$ game is given by \cite{Nowak}
\BE
\Pi_A(n)&=\frac{n-1}{N-1}a+\frac{N-n}{N-1}b,\nonumber \\
\Pi_B(n)&=\frac{n}{N-1}c+\frac{N-n-1}{N-1}d,
\EE
if the system is in state $n$. The parameters $a,b,c$ and $d$ specify the underlying game. We study the evolutionary dynamics defined by the birth-death process with rates \cite{TraulsenClaussen}
\BE
T^+_n&=\frac{1-u}{2}\left[1+\beta\Delta\Pi(n)\right]\frac{n(N-n)}{N^2}+\frac{(N-n)^2}{N^2}\frac{\mu}{2},\nonumber \\
T^-_n&=\frac{1-u}{2}\left[1-\beta\Delta\Pi(n)\right]\frac{n(N-n)}{N^2}+\frac{n^2}{N^2}\frac{\mu}{2},\label{eq:rates}
\EE
where $\Delta\Pi(n)=\Pi_A(n)-\Pi_B(n)$, and where $T^+_n$ is the rate at which a individual of type $B$ is replaced by an individual of type $A$ in state $n$, $T^-_n$ is the rate of the opposite event.
The parameter $\mu\in[0,1]$ represents the mutation rate, $\beta\geq 0$ is the intensity of selection.
The probability $P_t(n)$ of finding the system in state $n$ at time $t$  then obeys the master equation
\be
\label{eq:master}
\dot{P}_t(n)=T^+_{n-1}P_t(n-1)+T^-_{n+1}P_t(n+1)-(T^+_n+T^-_n)P_t(n).
\ee
We will denote the stationary distribution at asymptotic times by $\psi^*(n)=\lim_{t\to\infty} P_t(n)$. Following \cite{Levin}, the mixing time $t_{\text{mix}}(\varepsilon)$ is defined as $t_{\text{mix}}(\varepsilon)=\text{min}\{t:d(t)\leq\varepsilon\}$ where the variational distance $d(t)=\frac{1}{2}\sum_{n}\left|P_t(n)-\psi^{\ast}(n)\right|$
measures the distance of the probability distribution $P_t(\cdot)$ from the stationary distribution $\psi^{\ast}(\cdot)$ \cite{Levin}. 

In order to characterize the behaviour of mixing times in different scenarios, it is useful to first consider the limit of $N \to \infty$. In this case, the fraction of individuals of type $A$,  $x=n/N$ , fulfills the deterministic replicator-mutator equation (RME)
\BE
\dot{x} \!=\! \beta x (1-x)[(a\!-\!c)x-(d\!-\!b)(1-x)](1-\mu)+\tfrac{1-2x}{2}\mu.\label{eq:rme}
\EE
The number, position and stability of the fixed points of Eq. (\ref{eq:rme}) generally depend on the parameters $\beta$ and $u$, as well as on the underlying game \cite{Nowak, Bomze}.  

\begin{figure}[t!!]
\centering
\includegraphics[width=1\columnwidth]{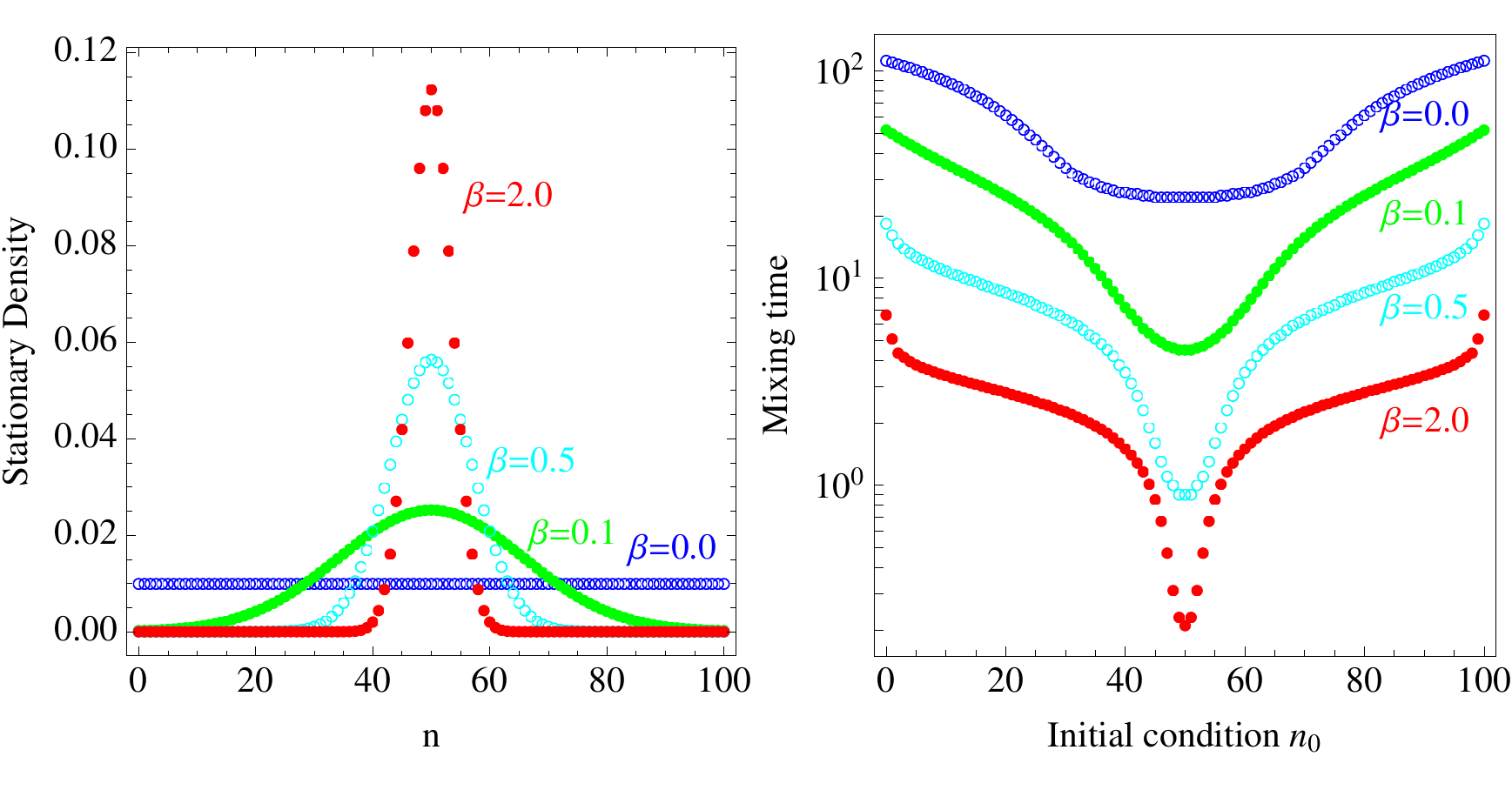}
\caption{(Colour online) Symmetric co-existence game in a population of $N=100$ individuals. Left: Stationary distributions of the stochastic dynamics (Eq. (\ref{eq:master})). Right: Mixing time ($\varepsilon=1/4$) when the stochastic process is started from a Dirac distribution at $P_0(n)=\delta(n-n_0)$. The mutation rate is $\mu=1/101$, leading to a uniform stationary distribution at $\beta=0$. \label{fig:anti}}
\end{figure}

We first study a symmetric coexistence game, defined by $a=1$, $b=2$, $c=2$, $d=1$. 
In this scenario, Eq.~(\ref{eq:rme}) has one stable fixed point at $x^*=1/2$, and no other fixed points.  The left panel of Fig.~\ref{fig:anti} shows the stationary distributions of the resulting Markov chain for different intensities of selection $\beta$. With increasing $\beta$, the distribution concentrates on the area around the deterministic fixed point. The corresponding mixing times are shown in the right-hand panel, starting the dynamics in a single state, $P_0(n)=\delta_{n,n_0}$, for varying $n_0$. 
Increasing the intensity of selection $\beta$ reduces the mixing time, 
because mixing is governed by the deterministic flux, which increases with $\beta$.
For $n_0 \neq \tfrac{N}{2}$, the mixing time is limited by this term.  Note that the fixation time in this game increases exponentially with $N$, as motion against the deterministic flux is required \cite{Antal}.

\begin{figure}[t!!]
\centering
\includegraphics[width=1\columnwidth]{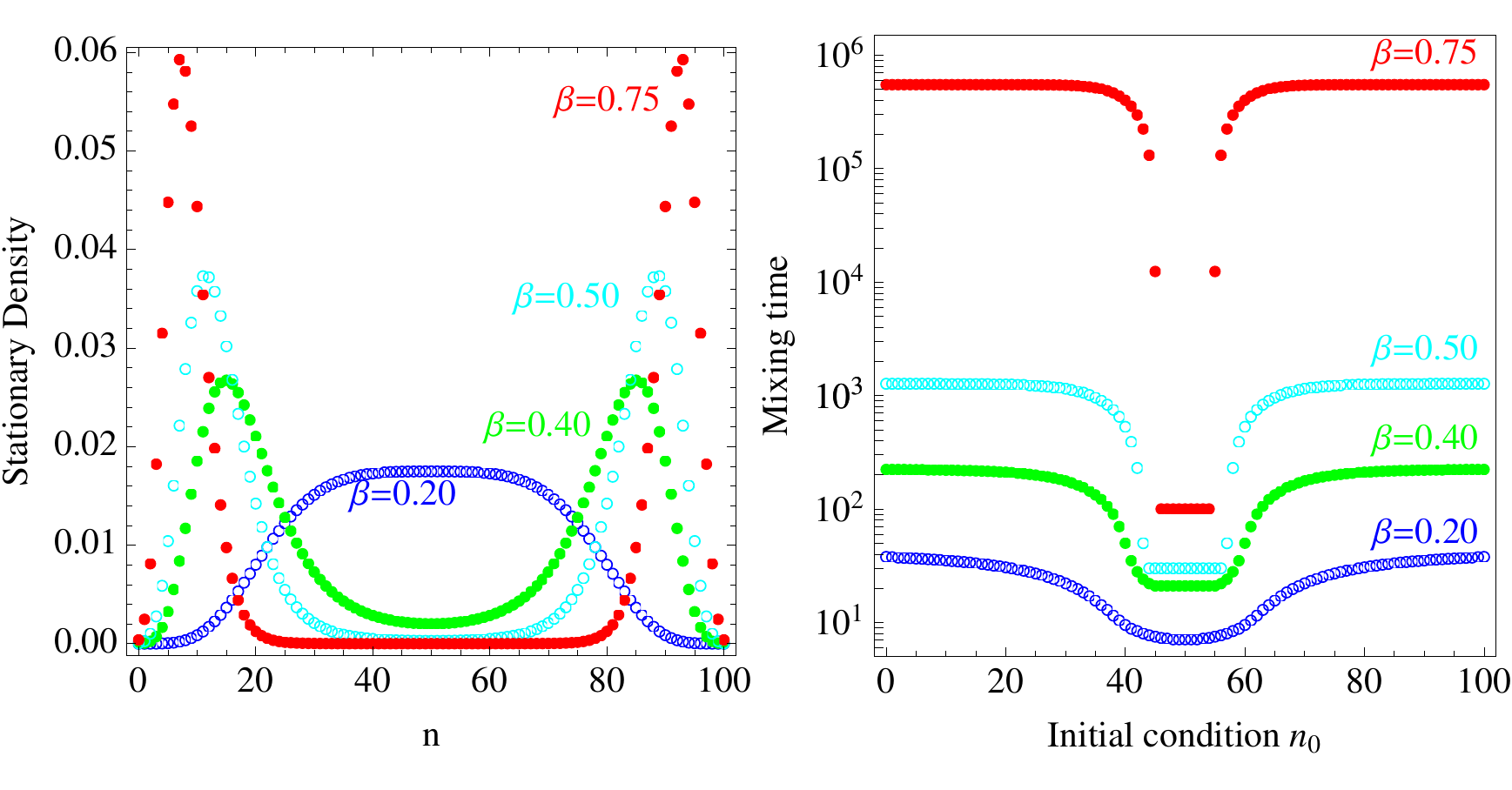}
\caption{(Colour online) Symmetric coordination game in a population of $N=100$ individuals. Left: Stationary distribution of the master equation (\ref{eq:master}).
Right: Mixing time ($\varepsilon=1/4$) as a function the position of the Dirac distribution from which the dynamics is started. The mutation rate is $\mu=1/101$, leading to a flat stationary distribution for $\beta=0$.}
\label{fig:coord}
\end{figure}

Next, we address a symmetric coordination game with parameters $a=2$, $b=1$, $c=1$, $d=2$. 
In this case, Eq.~(\ref{eq:rme}) has an unstable fixed point at $x_1=\tfrac{1}{2}$ and two stable fixed points $x_2$ and $x_3$ near $x=0$ and $x=1$ respectively. %
Fig.\ \ref{fig:coord} shows the bimodal stationary distributions and mixing times for this game. 
With increasing $\beta$, the distribution becomes sharply peaked around the two stable fixed points. The mixing time is then very sensitive to the initial condition. 
When the system is started near to either of the stable fixed points, the system will quickly tend to a quasi-stationary distribution (QSD) around one of the fixed points. 
The probability will then slowly leak over to the other side on a time scale exponentially slow in $N$. 
Thus, the mixing times increase with increasing intensity of selection $\beta$, whereas fixation times would decrease with $\beta$ \cite{Antal}. 
We can exploit this separation of time scales to calculate the mixing time analytically, similar to the problem of calculating the mean switching time between quasi-stationary states.

Let us assume that we start to the left of the unstable fixed point. The time-dependent ansatz we use for the probability distribution is
\beq
\label{eq:qsdapprox}
\psi^{{\rm leak}}(n)= \left\{ 
\begin{array}{ll}
 \psi^{\ast}(n)(1+e^{-E t}) & \quad n<n_1 \\ 
 \psi^{\ast}(n)(1-e^{-E t}) & \quad  n>n_1
\end{array}\right.,
\eeq
where $-E<0$ is the eigenvalue of the slowest decaying mode of the problem, and where $n_1$ corresponds to the central unstable fixed point. This ansatz is valid on an exponentially long time scale in $N$. Explicitly calculating the variational distance between the ansatz of Eq.~\eqref{eq:qsdapprox} and the stationary distribution, we find $t_{\mbox{\tiny mix}}(\varepsilon)=- E^{-1}\ln[2\varepsilon]$ 
such that the problem reduces to finding the eigenvalue $-E$. Based on Eq.~\eqref{eq:qsdapprox}, the current through the central fixed point, is given by $J(t)=\frac{d}{dt}\sum_{n<n_1} \psi^{\rm leak}(n)=-(E/2)e^{-E t}$.
Thus we can find $E$ from the initial current, $J(0)=-E/2$. 
Calculating the mixing time then reduces to the well studied problem of determining the escape current in a bistable potential \cite{Hanggi1990}. 

As this is a one-step process, exact expressions exist for the mean first passage times \cite{Gar03}. One avenue would be to derive the large $N$ asymptotics for these \cite{Doering2007,Antal,Redner}. We do not follow this approach here, instead we calculate the initial current through the unstable fixed point based on the celebrated WKB approximation. 
This has two advantages: first, this method is valid for a much wider range of problems, such as those with multiple jumps \cite{Escudero2009,Assaf2010}, or of higher dimensions \cite{Dykman1994}. 
Secondly, the stationary distribution $\psi^{\ast}(n)$ is calculated as a by-product. Our approach also complements recent studies which have successfully introduced these methods to evolutionary game theory by calculating fixation times in evolutionary games without mutation \cite{Mobilia2010}. 
Before we describe the main steps of the calculation, it is useful to introduce $x=n/N$ and to expand the transition rates, 
Eq.~\eqref{eq:rates}, into powers of $1/N$ in leading and sub-leading order. 
Specifically, we write $T^{\pm}(Nx)=w_{\pm}(x)+u_{\pm}(x)/N$ \cite{Assaf2010}. We use the WKB ansatz,
\beq
\label{eq:wkb_ansatz}
\psi^{{\rm WKB}}(x)=\exp[-NS(x)-S_1(x)],
\eeq
where both $S(x)$ and $S_1(x)$ are assumed to be of order $N^0$.  It is important to note the difference between ansatz \eqref{eq:qsdapprox} and \eqref{eq:wkb_ansatz}.
Ansatz \eqref{eq:qsdapprox} is valid on exponentially long time scales and about the asymptotes of the distribution and it takes account of the back current through the central fixed point. Ansatz \eqref{eq:wkb_ansatz} is valid everywhere, but only on short time scales. It can therefore be used to calculate the initial current. 
We proceed by inserting Eq. \eqref{eq:wkb_ansatz} into Eq. \eqref{eq:master}, assuming (quasi-) stationarity ($\partial_t \psi^{\rm WKB}=0$) and expanding the resulting equation into powers of $N^{-1}$ \cite{Bender,Escudero2009,Assaf2010,Mobilia2010}. In lowest order,  we find a Hamilton-Jacobi equation,
\beq
H(x,p)\equiv w_+(x)(e^{p}-1)+w_-(x)(e^{-p}-1)=0,
\label{eq:ham}
\eeq
where $p=\partial_x S$. This constitutes an equation for $S(x)$ and it has two solutions: (i) the activation solution 
\be\label{eq:s}
S(x)=\int^xd\xi \ln \left[ \frac{w_+(\xi)}{w_-(\xi)} \right],
\ee
 and (ii) the so-called relaxation solution $S(x)=0$. 
In next order, we find the activation solution
\beq\label{eq:s1}
S_1(x)=\int^x d\xi \left( \frac{u_-(\xi)}{w_-(\xi)}-\frac{u_+(\xi)}{w_+(\xi)}\right) +\frac{1}{2}\ln[w_+(x) w_-(x)],
\eeq
and the relaxation solution $S_1(x)=\ln\left[H_p(x,0)\right]$, where $H_p=\partial H/\partial p$. In our setup the activation solution describes the behaviour of the QSD to the left of $x_1=1/2$. 
The relaxation mode, describing deterministic motion to the right of $x_1$, will play no significant role in our further analysis.

In order to complete the calculation two main tasks remain: (a) the activation solution $\psi^{\rm WKB}$ defined by Eqs. (\ref{eq:wkb_ansatz},\ref{eq:s},\ref{eq:s1}) needs to be normalized, and (b) the QSD needs to be connected to the initial current $J(0)$ through $x_1$. These tasks can be addressed by performing a Kramers-Moyal expansion of the master equation (\ref{eq:master}) around the unstable fixed point $x_1$. Writing $f_\pm(x)=Nw_\pm(x)\psi(x)$ we find 
\be
\partial_x\left[\sum_{r=\pm 1} \frac{r}{N}f_r(x)-\frac{r^2}{2N^2}f_r'(x)\right]=0,
\ee
where the term in the square bracket is identified as the divergence-free probability current $J(0)$. Further algebraic manipulations then lead to \cite{Escudero2009},
\beq
J=\psi^{\rm WKB}(x)H_{px}(x,0)(x-x_1)-\frac{H_{pp}(x,0) \partial_x \psi^{\rm WKB}(x)  }{2N},
\label{eq:kramer}
\eeq
where $H_{pp}=\partial^2 H/\partial p^2=w_++w_-$ and $H_{px}=\partial^2 H/\partial x\partial p=w_-w_+'/w_+-w_+w_-'/w_-$. Re-arranging Eq. (\ref{eq:kramer}) one has
\beq
\psi^{\rm WKB}(y)=\frac{J\sqrt{\pi}}{H_{px}\ell}e^{y^2}\erfc(y), 
\quad \hbox{where} \quad 
y= \frac{x-x_1}{\sqrt{H_{pp}/(NH_{px})}}.
\eeq
The final step then consists in matching the asymptote, $\psi^{\rm WKB}(y)=J\sqrt{\pi}e^{y^2}$, valid for $y\ll -1$, with the relaxation-mode solution $\psi^{\rm WKB}(x)=Ae^{-NS(x)-S_1(x)}$ (with $S(x)$ and $S_1(x)$ given by Eqs.~(\ref{eq:s}) and (\ref{eq:s1})). The normalisation constant $A$ is obtained from a Gaussian approximation of the relaxation solution about the fixed point $x_2$ \cite{Dykman1994}.
 
Carrying out this procedure the initial current is found to be exponential in $N$: 
\BE
J(0)&=&\frac{H_{pp}(x_1)}{4\pi N}\sqrt{S''(x_2)|S''(x_1)|}  \nonumber \\
&&\times e^{N[S(x_2)-S(x_1)]+S_1(x_2)-S(x_1)}.
\EE
Finally the mixing time is 
\beq
t_{\text{mix}}(\varepsilon)=\ln(2\varepsilon)/(2J(0)).
\label{eq:mix_time}
\eeq

\begin{figure}[t!!]
\centering
\includegraphics[width=0.95\columnwidth]{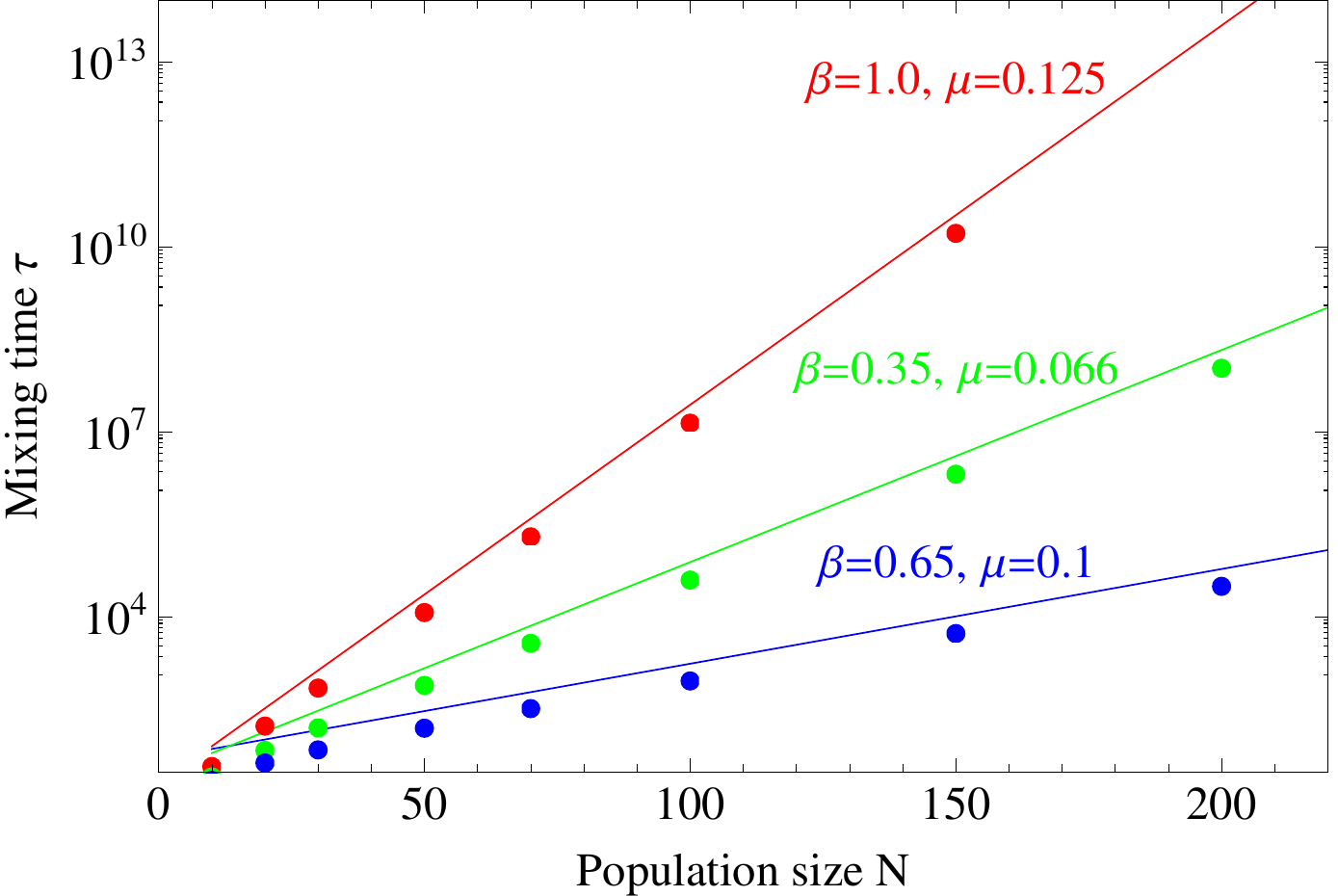}
\caption{(Colour online) Mixing time ($\varepsilon=1/4$) for the symmetric coordination game as a function of the population size $N$. Lines are calcuated from Eq. (\ref{eq:mix_time}), dots are from integration of the master equation.}
\label{fig3}
\end{figure}

Figure \ref{fig3} shows the mixing times calculated via the WKB method along with direct computation from a numerical integration of the master equation. Agreement is generally very good, except for small values of $N$ when the introduction of continuous variable $x$ as well as the expansion in powers of $N^{-1}$ become inaccurate. The slight offset between the two sets of results is due to the error introduced in the calculation by assuming Gaussian approximation made when normalizing the QSD $\psi^{{\rm WKB}}(\cdot)$. Better agreement could be obtained by normalizing the distribution exactly. 
However, this requires a numerical approach, whereas our final result is more explicit.  

While the WKB approach can successfully be employed to obtain mixing times, there are limitations to this method. One potential problem is the divergence of the WKB solution $\psi^{\rm WKB}(x)$ at the boundaries of the system.
This does not affect the outcome of our calculation as long as the stable fixed points of the RME are not too close to the boundaries of phase space, but it does limit the range of $\beta$ and $u$ for which it is valid. 
We stress that the ansatz \eqref{eq:qsdapprox} still applies, but the eigenvalue $E$ needs to be calculated via a different approach. The methods we have presented lend themselves to generalisation.  For example the assumption of symmetry of the problem can be relaxed, the unstable fixed point need not be at $x=1/2$. 

In summary, we have introduced the concept of mixing times for evolutionary dynamics with mutation. As intensity of selection is increased, the mixing times in co-existence games decrease. In coordination games, one observes the opposite trend. In both cases the behaviour of mixing times is opposite to that of fixation times in the corresponding systems without mutation. The concept of mixing times may often be more appropriate for many biological systems than the computation of fixation times, in particular when effects of mutation or immigration cannot be ignored \cite{Eriksson}. As shown in our work, tools from theoretical physics can be used to successfully estimate mixing times based on semi-analytical considerations. We expect this to be useful not only for biological systems, but also for models of social dynamics and other interacting many-particle processes.

{\em Acknowledgments.} AJB acknowledges support from the EPSRC and the ARC Discovery Projects funding scheme (project number DP110102893). TG is supported by RCUK (reference EP/E500048/1), and by the EPSRC (references EP/I005765/1 and EP/I019200/1).

\end{document}